# The role of thermal fluctuations and vibrational entropy for the δ-to-α transition in hybrid organic-inorganic perovskites: the FAPbI$_3$ case

Junwen Yin[1,2], Gilberto Teobaldi,[3,4] Li-Min Liu[2, *]

1. Beijing Computational Science Research Center, Beijing 100193, China
2. School of Physics, Beihang University, Beijing, 100083, China
3. Scientific Computing Department, STFC UKRI, Rutherford Appleton Laboratory, Harwell Campus, OX11 0QX Didcot, United Kingdom
4. School of Chemistry, University of Southampton, Highfield, SO17 1BJ Southampton, United Kingdom

\* liminliu@buaa.edu.cn

## Keywords

## Abstract

Formamidinium lead halide (FAPbI$_3$), as a typical hybrid organic-inorganic perovskite, has attracted considerable interest due to its band gap suitable for visible light absorption and good thermal stability. A barrier to the use of FAPbI$_3$ in commercial, stable devices is its unwanted black-to-yellow (non-perovskite to perovskite, commonly known as δ-to-α) phase transition at around 300 K. The intrinsic mechanisms of such phase transition are far from clear, being the detailed structural description for the α-phase still missing. By combined Density Functional Theory (DFT) calculations, lattice dynamics analysis and DFT molecular dynamics simulations, we assign the α-phase to the highly dynamic tetragonal phase, with the high-symmetry cubic structure emerging as a dynamically unstable maximum in the system's potential energy landscape. We demonstrate computationally that the diffraction-observed cubic structure is the result of the averaging of different tetragonal distortions sampled in the experimental detection time scale as a result of the enhanced FA dynamics, instead of a static system of cubic symmetry. Further finite-temperature Gibbs free energy calculations confirm that the δ-to-α transition should be considered as a hexagonal-to-tetragonal transition in contrast to the previous hexagonal-to-cubic assignment. More importantly, the simulations indicate that the driving force of the process is the vibrational entropy difference rather than the rotational entropy as previously proposed. These results point out the dynamical nature of the α-phase, the importance of the overlooked tetragonal structure, and the key role of the vibrational entropy in perovskite-related phase transitions, the harnessing of which is critical for successful uptake of ABX$_3$ hybrid organic-inorganic perovskites in commercial applications.

# Introduction

Materials with perovskite crystal structures are well-known to present complex phase-transition landscapes due to the temperature- and pressure-dependent competition of different distortions from the ideal, cubic lattice of the system.[1–3] These phase transitions are invariably accompanied by changes in physicochemical properties e.g., electronic structure as well as optical and magnetic features that are vital for technological applications.[4,5] It follows that understanding the atomistic mechanisms of phase transitions in perovskite-based functional materials is crucial for their rational function-tailored design and fabrication.[6–12] Among numerous perovskite materials, in the last decade hybrid organometal halide perovskites have received significant attention for solar cells applications due to rapid improvements of their efficiency. With a certified record of 25.2%, the efficiency of hybrid perovskite solar cells has started to rival that of conventional silicon-based technologies.[13–18] Due to their inexpensive solution-based fabrication and high-power conversion efficiency, hybrid perovskites have been receiving growing attention for applications in different fields ranging from photovoltaics, optoelectronics and sensing devices.[19–22] Their appealing performance and efficiency is understood to be due to a combination of a suitable band gap and electronic structure properties starting from a low exciton binding energy.[23,24]

As one of the most popular materials among ternary $ABX_3$ perovskite materials, $FAPbI_3$ ($CH(NH_2)_2PbI_3$) exhibits a high light absorption coefficient and long charge-carrier diffusion lengths.[25,26] Its complex structural transition between the non-perovskite (hexagonal) and perovskite (orthorhombic, tetragonal, cubic) phases, which is usually referred to as δ-to-α transition (see Figure 1), has been studied using neutron powder diffraction (NPD[27–29]). The complex phase transitions between different phases are understood to be dominated by the change in temperature.[30] In most cases, the α phase turns out to be stable at high temperatures whereas the δ phase is preferred at room temperature. The phase transition temperature between the α phase and the δ one ranges from 290 K to 400 K, depending on the fabrication method, heating or cooling of the sample, and the working environment.[31,32] Although numerous efforts have been devoted to stabilizing the α phase, the atomistic mechanisms of the δ-to-α transition remain not fully understood. Previous reports demonstrated the importance of the rotational entropy of the $FA^+$ cations for the hexagonal-to-cubic, δ-to-α phase transition using neutron diffraction and first-principles calculations.[32] However, the potential role of the vibrational entropy and of the metastable tetragonal phase for the mechanism of the δ-to-α transition have so far remained overlooked.

To identify the δ-to-α transition mechanism, a detailed structural description of the α phase is necessary. Although great efforts have been made by multiple research groups, such a subject remains controversial. Weller *et al.* assigned the α phase to a cubic structure with *Pm-3m* space group using high resolution NPD, while Stoumpos *et al.* concluded that the α phase exhibits a trigonal structure with a *P3m1* space group using single crystal X-ray diffraction analysis.[27,33] Zeng *et al.* assumed that the α black phase is likely to be constituted by micro-domains of cubic or distorted cubic structures.[34] Previous research in inorganic perovskites claimed the instability of cubic structures by calculating their complex energy landscapes.[35,36] A combined temperature resolved UV-vis absorption and ab initio simulations in MAPbI$_3$ (MA=CH3NH3) by Angelis *et al.* suggested that the high-temperature phase of MAPbI$_3$ seldom exhibits a cubic structure, rather a distorted tetragonal structure.[37,38] These results raise legitimate questions as to whether the α phase of FAPbI$_3$ may not actually correspond to a cubic structure as currently accepted by most researchers in the field.

In order to both revisit the atomic structure of the α phase and identify the driving force of the δ-to-α transition, we have carried out lattice dynamics analysis and DFT molecular dynamics (MD) simulations. The results reveal that due to the unique perovskite structures sustained by vibrational anharmonicity instead of covalent or ionic interactions, the α phase is a highly dynamical system that deviates from a high-symmetry cubic structure when sampled over time at temperatures above 0 K. The results suggest that averaging of numerous tetragonal symmetries sampled during the macroscopic timescale of acquisition for diffraction data is responsible for the (incorrect) assignment of the α phase to a cubic structure. Combining the DFT and lattice dynamics results, the vibrational entropy difference between the hexagonal and tetragonal phases at 290 K can compensate for the difference in internal energy between these two systems. Our results indicate that the FA$^+$ cations in both phases exhibit similar dynamics, ruling out earlier assumptions of a major role for the FA$^+$ rotational entropy in the δ-to-α transition.[32] Based on these results, we conclude that δ-to-α transition should be considered as a hexagonal-to-tetragonal transition, and that the vibrational entropy difference is the dominant driving force of such a transition. These findings allow reconciliation of the long-standing contradiction of a cubic symmetry observed for the α-phase and the imaginary frequencies (dynamic instability) computed at different levels of theory for cubic perovskite structures.

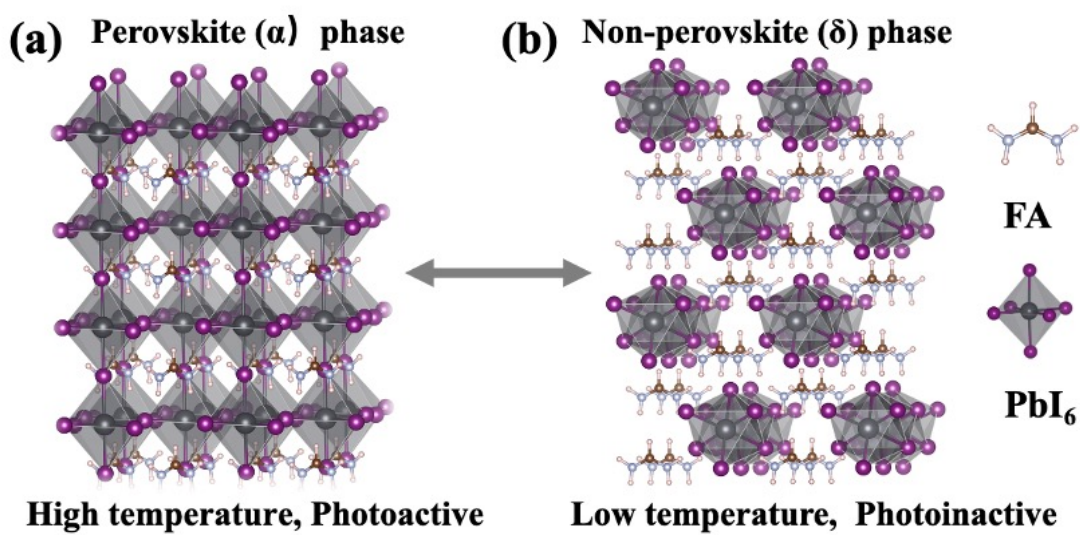

**Figure 1. Phase transition diagram of FAPbI$_3$.** (a) The perovskite phase (α) of FAPbI$_3$, which is the photoactive phase, exists at high temperatures. (b) The non-perovskite

phase (δ) of FAPbI$_3$, which is the photoinactive phase, exists at low temperatures. A temperature triggered phase transition takes place between these two phases.

## Results and discussion

### *Geometric identification of FAPbI$_3$*

For ternary ABX$_3$ perovskite materials, the A site cation usually resides in the voids of the BX$_3$ octahedral framework. The displacement of the A cation and the tilting of the BX$_3$ octahedral network account for the complex structural landscapes of the ABX$_3$ structures as a function of the external temperature, pressure and applied electric field.[6,29] FAPbI$_3$ has an ABX$_3$ structure in which the A, B, X sites are FA$^+$, Pb and I, respectively. As shown in Figure 2 (structures 1-4), in perovskite FAPbI$_3$, the FA$^+$ cations reside in the cubo-octahedral cavities of the corner-sharing lead halide octahedral network and are hydrogen-bonded to the PbI$_3$ octahedrons. Conversely, in the non-perovskite hexagonal phase, the PbI$_3$ units adopt a face-sharing (not corner-sharing) geometry, as shown in Figure 2 (structures 5-6). To obtain the ground state of the cubic, tetragonal, and hexagonal structures of FAPbI$_3$, different ferroelectric (FE) and antiferroelectric (AFE) distribution of the FA$^+$ cations in each phase are optimized and their energy compared. As shown in Figure 3, structures 1, 3, 5 display FE-ordering, while systems 2, 4, 6 have AFE-ordering. Comparison of the DFT calculated internal energies (Table S1) reveals negligible differences between the FE and AFE ordering for the cubic (0.03 eV per unit cell) ad hexagonal (0 eV per unit cell) structures. Conversely, in the tetragonal phase, the FE-ordering is favored by 0.5 eV per unit cell over the AFE one. For sake of clarity, in the following we present the discussion of the atomic structures 1, 3, 5 (FE-ordering) for the three different structures separately.

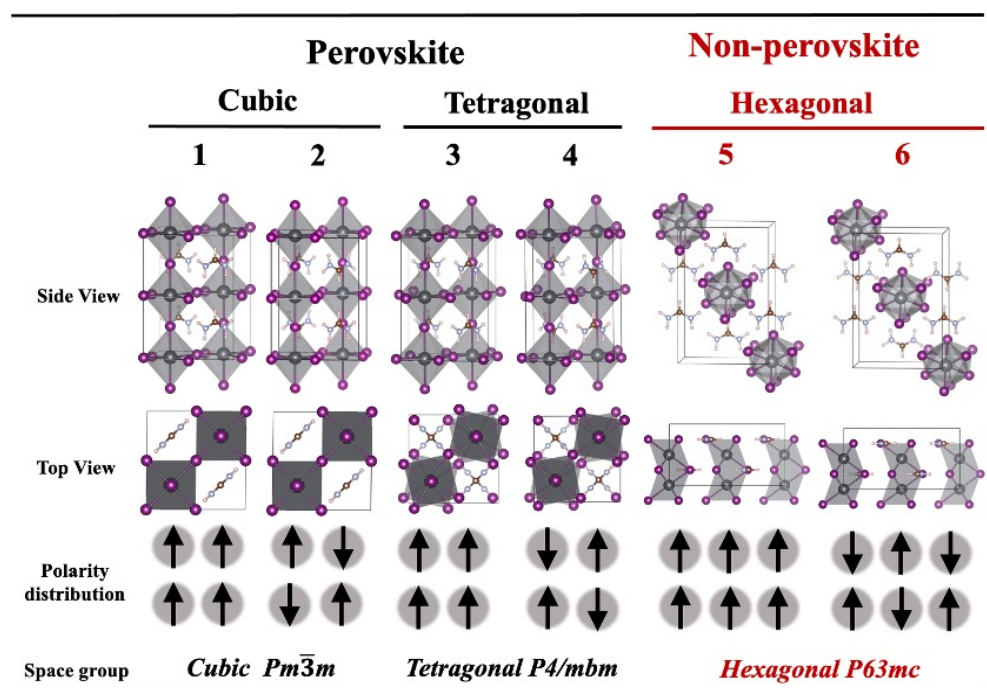

**Figure 2. Computational models used in the DFT calculations.** Structural representation of the three different structures (cubic, tetragonal, and hexagonal) of FAPbI$_3$ with the different possible distributions for the FA$^+$ cations and associated electrostatic dipoles. For the two perovskite structures (cubic and tetragonal), the octahedral tilting is different. For the non-perovskite hexagonal phase, there is no octahedral tilting. The bottom panel summarizes the polarity distribution for the two possible FE- (structures 1, 3, 5) and AFE- (2, 4, 6) orderings of the FA$^+$ electrostatic dipoles in each phase. The purple, gray, white, brown, and violet balls stand for the I, Pb, H, C and N atoms, respectively.

*Cubic.* The cubic phase is the high-temperature structure of FAPbI$_3$, which is observed over around 300 K. High resolution NPD results assign the Pm$\bar{3}$m space group to this high-temperature structure.[27] Due to the isotropic rotational behavior and random orientation of the FA$^+$ cations at high temperature, the lattice parameter of cubic FAPbI$_3$, tends to be more isotropic than for the other phases. As shown in Table 1, due to the dipole alignment along the z axis for the FE ordering, structure 1 has primitive-cell lattice parameters of a = b = 6.39 Å, c = 6.27 Å (transformed from the 48-atom cell lattice parameter in Table 1). Previously reported experimental results for the lattice parameters are a = b = c = 6.3855 Å.[27] The underestimation of the lattice- parameter should be attributed to the fixed FA$^+$ cations and the neglect of temperature effects. According to Glazer notation, the octahedral tilting pattern for the cubic phase is $a^0b^0c^0$.[42,43]

**Table 1. First-principles calculations results.** Unit cell (48 atoms, 4 formula units) relative energies and lattice parameters for the three different perovskite and non-perovskite structures of FAPbI$_3$. These correspond to structures 1, 3, and 5 in Figure 2, respectively.

|  |  | Perovskite | | Non-perovskite |
|---|---|---|---|---|
|  |  | **Cubic** | **Tetragonal** | **Hexagonal** |
| **Energy (eV)** |  | 1.04 | 0.3 | 0 |
|  | a | 9.05 | 8.85 | 9.50 |
| **Lattice parameter (Å)** | b | 9.05 | 8.86 | 7.61 |
|  | c | 12.54 | 12.62 | 14.32 |

*Tetragonal.* The tetragonal FAPbI$_3$ is observed for temperatures lower than 285 K as a metastable phase.[44] Recent NPD analysis assigns the P4/mbm space group to the tetragonal phase.[44] The octahedral tilting pattern of the tetragonal phase is $a^0a^0c^-$ in Glazer notation. For this phase, the rotational dynamics of the FA$^+$ cations are slightly inhibited due to the lower temperature.[30] The less isotropic orientation of the FA$^+$ cations in this structure leads to distortion of the PbI$_6$ octahedrons. These distortions further stabilize the structure, as reflected by the energy gain (-0.74 eV per unit cell) with respect to the cubic phase (structures 1 and 3 in Figure 2 and Table 1).

*Hexagonal.* The hexagonal phase is the preferred phase of FAPbI$_3$ at room temperature. This is confirmed by its lowest DFT energy obtained in Table 1. FAPbI$_3$ undergoes a first-order phase transition from the perovskite phase to the non-perovskite, hexagonal phase at around 300 K. Different from the corner-sharing octahedrons of the perovskite structure, the PbI$_6$ unit adopt a face-sharing geometry. These changes in the octahedral framework make the Glazer notation no longer applicable. The significantly lower energy of the hexagonal phase compared to that of the cubic system indicates that the substantial reduction of the system's symmetry in the hexagonal phase is energetically advantageous.

From the calculated DFT-energies in Table 1, it can be concluded that, at 0 K, the hexagonal phase is the ground state structure for FAPbI$_3$ due to its lowest internal energy: 0.3 eV and 1.04 eV per unit cell lower than that of the tetragonal and cubic phases, respectively. However, the 0 K DFT geometry optimization results cannot explain the temperature dependent interconversion between these three phases as well as the huge internal energy difference between the cubic structure and the other two. To this end, we next turn to lattice dynamics calculations and phonon properties analysis.

***Phonon dispersion and lattice dynamics analysis***

Following the previous analysis of the structure and relative energy of the cubic, tetragonal, and hexagonal structures of FAPbI$_3$, we next turn to their phonon dispersion calculated at DFT level within the harmonic approximation. For this analysis, we focus on the energy-favored structures (1, 3, 5 in Figure 2 and Table 1) for each considered phase. Figure 3 reports the phonon density of states projected onto the FA$^+$ cations and PbI$_3$ octahedrons for the cubic, tetragonal, and hexagonal structures. The phonon band structures are shown in Figure S1. It can be seen in Figure 3 that the three FAPbI$_3$ phases have similar phonon density of states with three characteristic energy regions: (1) a low-frequency band from 0 to 6 THz; (2) a mid-frequency band from 10 to 55 THz; (3) a high-frequency band over 100 Thz.

For N=12 atoms in the primitive cell, FAPbI$_3$ presents 3N=36 vibrational eigenmodes in total. Of these 36, 12 (3N$_1$, with N$_1$ = 4: the number of atoms in PbI$_3$) modes belong to the PbI$_3$ fragment, and 24 (3N$_2$, with N$_2$ = 8: the numbers of atoms in FA$^+$) to FA$^+$=CH(NH$_2$)$_2$$^+$ cations. The 12 PbI$_3$ modes include 9 vibrational modes and 3 acoustic translational modes of the octahedral lattice. The 24 FA$^+$ modes comprise 18 vibrational modes, 3 translational modes, and 3 rotational modes. Due to the large atomic mass difference of the PbI$_3$ octahedrons and FA$^+$ cations, as well as the difference between the ionic bonding of PbI$_3$ and the covalent bonding of the FA$^+$ cations, it is expected that the low-frequency band (0-6 THz) originates from the motion of PbI$_3$, while the mid- (10-55 THz) and high (>100 THz) frequency bands are associated with the dynamics of the FA$^+$ cations. However, in Figure 3 one can observe an overlap of the PbI$_3$ and FA$^+$ modes in the low-frequency band, which indicates vibrational coupling between the two fragments. Normally, the 6 translation and rotational modes are not



reflected in the vibrational spectrum for an isolated nonlinear molecule. However, for the FA$^+$ cations that reside in a cubo-octahedral cavity, this is not the case. These 6 (translational and rotational) modes are strongly coupled to the 9 vibrational modes of PbI$_3$, as demonstrated by the overlap between the FA$^+$- and PbI$_3$- decomposed density of states in the low frequency region (< 6 THz in Figure 3). The 9 modes of PbI$_3$ can be attributed to the stretching of the Pb-I bonds and breathing of the octahedrons. The 6 modes of the FA$^+$ cation in the low-frequency region involve the pivoting motion of the fragment in its hosting cavity. The mid- and high- frequency bands above 10 THz comprise exclusively vibrational modes of the FA$^+$ cation, specifically molecular twisting, angle-bending, and bond-stretching motions.

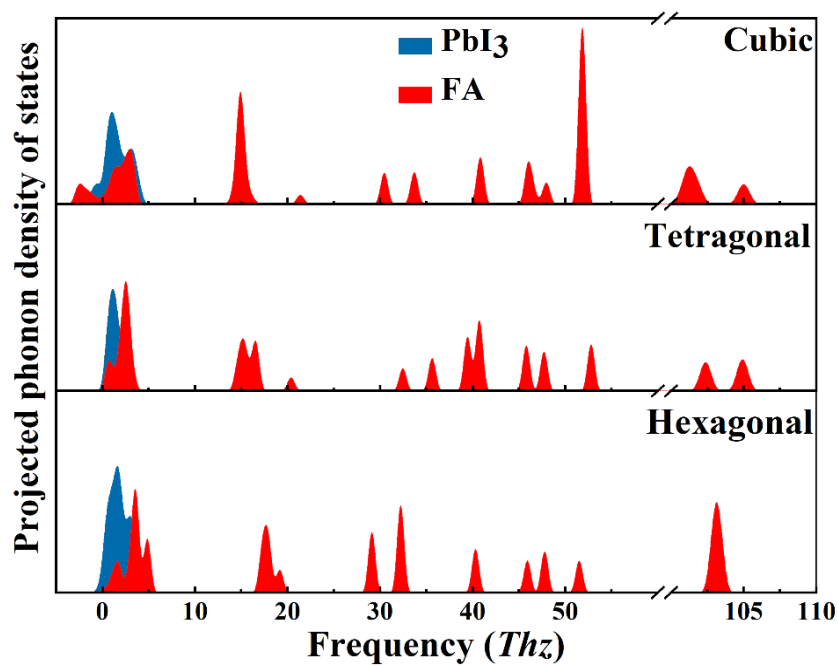

**Figure 3.** FA- and PbI$_3$-decomposed phonon density of states for the cubic, tetragonal, and hexagonal structures of FAPbI$_3$. The blue and red traces represent the vibrations of the PbI$_3$ octahedrons, and FA$^+$ cations, respectively.

Inspection of both Figure 3 and Figure S1 readily reveals the presence of imaginary soft phonon modes of negative frequency at the boundary of the Brillouin zone for the cubic structure. Similar imaginary modes have been reported also for cubic MAPbI$_3$ by Walsh and coworkers.[6] These negative-frequency modes, linked to tilting of the PbI$_3$ octahedrons and molecular rotations, indicate dynamic instability of the cubic structure. This result demonstrates that in order to tackle the mechanisms of the δ-to-α transition, the dynamical characteristics of the high temperature α phase, beyond the static picture of the cubic structure used so far, must be taken into account. In the following we further explore the dynamical nature of the α phase by means of DFT-MD simulations.



*Temperature- and time-dependent structural evolution simulation*

We performed DFT-MD simulations for the α phase at 250 K and 400 K, starting from a high-symmetry cubic structure. To contain the computational costs of DFT-MD simulations within limits of practicability, we used a (304 atoms) 3 × 3 × 3 supercell. As evident even from qualitative visual inspection of the movies provided in the Supplementary Information, inclusion of the temperature in the simulations leads to distorted perovskite structures with tilted octahedrons and random orientations for the $FA^+$ cations throughout the equilibrated DFT-MD trajectories. On the basis of these results, it is to be concluded that the α-phase should be assigned to a tetragonal, rather than a cubic one. More specifically, combining the calculated energies for the optimized systems (Table 1), with the calculated phonon instabilities (Figure 3) and the results of the MD-DFT trajectories, the cubic structure turns out to be a local maximum in the energy landscape of the α-phase, with temperature-populated local minima being characterized by distorted tetragonal structures.

Figure 4a reports the calculated time evolution of the *running-average* atomic positions for the α structure at 250 K and 400 K. These are obtained by averaging the atomic structures along the DFT-MD trajectory (8 ps production time in total) over progressively larger time-windows. Clearly, the larger the time-window used the larger the numbers of structures included in the running-average results. As seen in Figure 4a, for reduced time-averaging intervals e.g., 1 ps at 400 K and 3 ps at 250 K, the average structure results to be tetragonal. However, as the sampling periodic increases, the time-averaged structures become cubic. In spite of their limited timespan and execution on inevitably finite atomistic models, these DFT-MD results demonstrate that time-averaging of the $FAPbI_3$ tetragonal distortions over sufficiently long times, as required for acquisition of detectable X-ray and neutron diffraction peaks, [27–29,33] leads to the appearance of an overall cubic structure as time-average of tetragonally distorted structures.

Analogous conclusions can be drawn by analyzing the time evolution of the I-Pb-Pb-I dihedral angle during the DFT-MD trajectory (Figure 4c). Whereas an undistorted cubic structure corresponds to I-Pb-Pb-I dihedral angles of 0°, tetragonal distortions result in changes of such angles from 0° as observed at both 250 K and 400 K. Notably, the occurrence of 0° I-Pb-Pb-I dihedral angles in the 400 K trajectory is 15% more frequent than at 250 K, indicating a higher prevalence of cubic-like structures (as average of different tetragonal distortions) as the temperature increases.



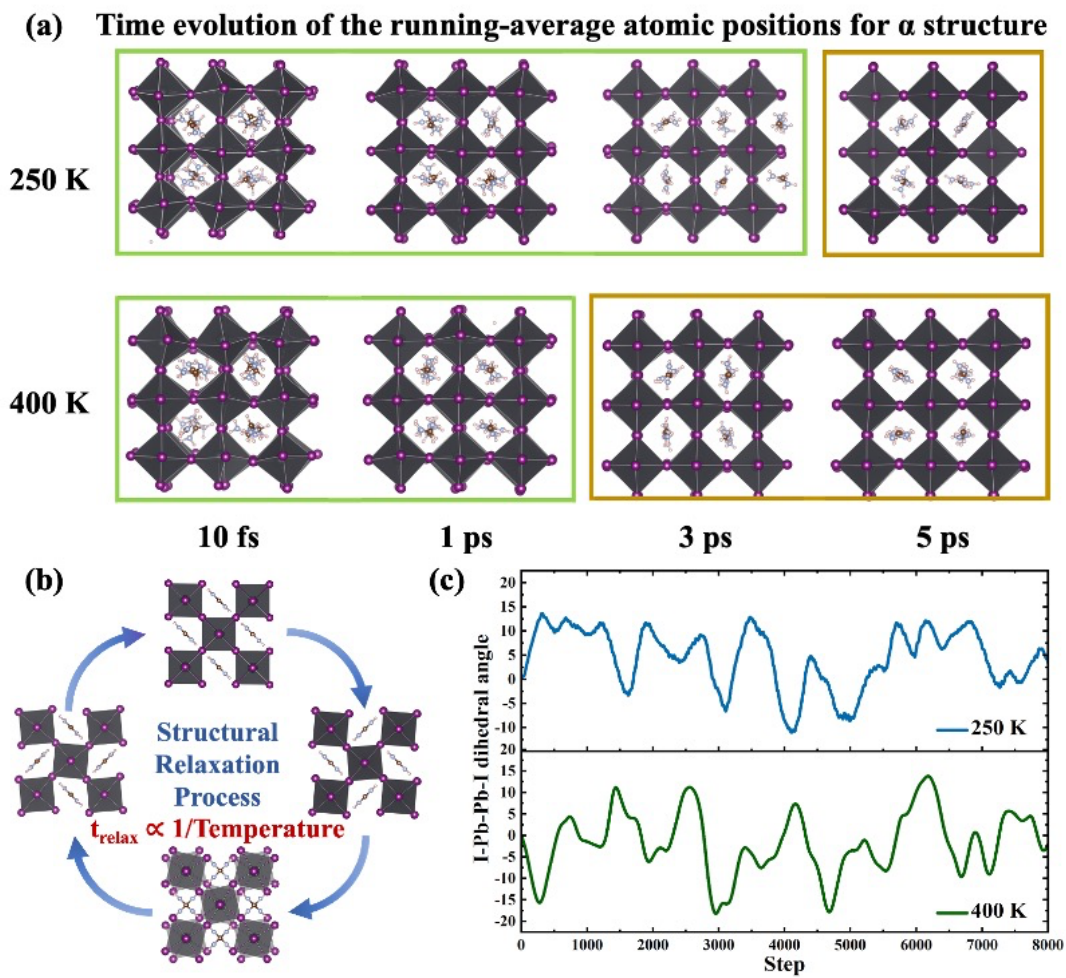

**Figure 4. DFT-MD simulations of the α phase at 250 K and 400 K.** (a) Time evolution of the *running-average* atomic positions for the α structure at 250 K and 400 K. The structures in the green frames correspond to tetragonal structures while those in yellow frames correspond to cubic structures. (b) Schematic representation of the dynamical behaviors of the α structure, which consists of an ensemble of differently distorted tetragonal structures. (c) Time evolution of the I-Pb-Pb-I dihedral angle in the α phase at 250 K and 400 K over an 8 ps time span.

Based on these results, the dynamical behavior of the α system at room temperature can be schematically summarized as in Figure 4b. The cubic structures appear as an instantaneous state which resides at a local maximum (Table 1 and Figure 3) of the potential energy surface of the α phase, with distorted tetragonal structures for the local minima on such a surface. Over relatively long times (relaxation time), the time evolution of the system sample several different tetragonal distortions, originating (diffracting as [27–29]) a time-averaged cubic structure. As time-averaging of less than 10 ps (Figure 4), way shorter than the acquisition time of powder diffraction techniques,[27–29] is needed for such a cubic average to appear, it is put forward that the α phase should be assigned to a dynamic tetragonal structure, rather than a cubic one as hitherto assumed. Based on this revised assignment for the atomic structure of the α phase, in



the following we explore the intrinsic mechanism of δ-to-α phase transition.

*Vibrational entropy and Gibbs free energy computation*
As previously discussed, $FAPbI_3$ exhibits a temperature-dependent δ-to-α phase transition. Given such temperature dependence, we next turn to investigating the role of entropy for these phase transitions. Under thermodynamic equilibrium, a finite temperature transition between different structures can be attributed to a difference in Gibbs free energy between the initial and final phases. In turn, the change in Gibbs free energy consists of changes in the internal energy (U) and vibrational/rotational Helmholtz free energy. Although Chen *et al.* have previously investigated the role of the rotational Helmholtz free energy for the transition between the cubic and hexagonal phases,[32] the combined role of vibrational free energy ($F_{ph}$) *and* vibrational entropy in the phase transition between different $FAPbI_3$ cubic, hexagonal, and tetragonal phases has remained overlooked. In the following we quantify and analyze these overlooked factors and their role for the temperature-dependent competition between perovskite α (cubic and tetragonal) and non-perovskite δ (hexagonal) phases of $FAPbI_3$.

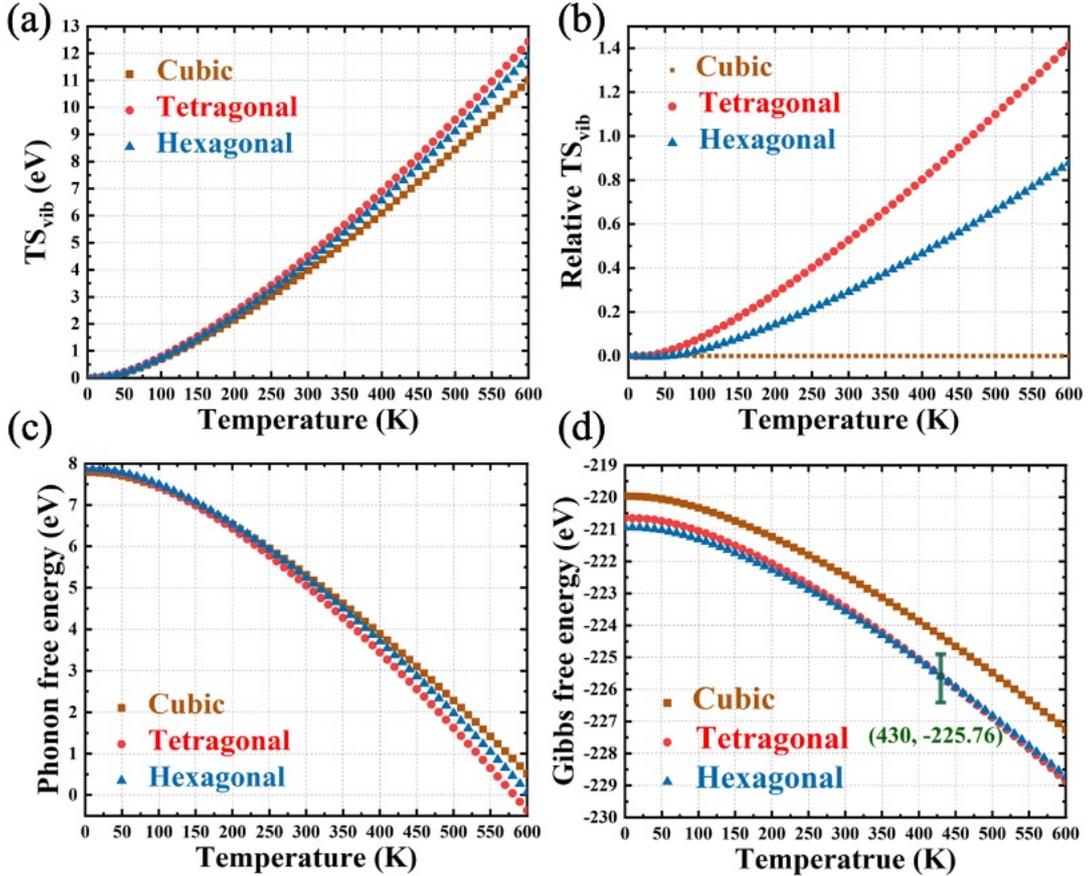

**Figure 5. Thermal properties of different structures by phonon calculation.** (a) Vibrational entropy contribution $TS_{vib}$, (b) $TS_{vib}$ of tetragonal and hexagonal structures relative to that of the cubic structure, (c) phonon free energies, (d) Gibbs free energies for the different $FAPbI_3$ structures. The green line marked point in panel (d) marks the calculated phase transition point (430 K, -225.76 eV).



As discussed above (Table 1) the relative internal energy (per unit cell) of the cubic, tetragonal, and hexagonal structures are 1.04, 0.3 and 0 eV at 0 K, which points to the hexagonal phase as the ground state phase at 0 K. As shown in Figure 5(c), the phonon free energy of the three phases decrease as the temperature increases. This is because an increasingly larger and negative vibrational entropy term (-$TS_{vib}$) is algebraically added to the (temperature independent) vibrational zero-point energy of the systems. As the vibrational entropy stabilizes a given phase by reducing its Gibbs free energy, the larger the vibrational entropy (the softer the vibrational modes of the systems), the larger the entropy stabilization for the given phase. To have a clearer view of the role of the vibrational entropy for the different phases, we plot their temperature dependences and ensuing differences in Figure 5(a) and Figure 5(b), respectively.

As seen in Figures 5(a-b), and regardless of the temperature, the cubic (tetragonal) structure systematically yields the smallest (largest) vibrational entropy $TS_{vib}$. The results for the hexagonal structure are intermediate and lie between the traces for the cubic and tetragonal phases. The larger vibrational entropy of the tetragonal structure with respect to the hexagonal one points to an inversion in the relative Gibbs free energy of the systems as the temperature is increased. From Figure 5(b), the difference in $TS_{vib}$ per unit cell between the tetragonal and hexagonal phases becomes ≥0.3 eV for temperatures T ≥ 290 K. This can over-compensate the 0.3 eV difference in internal energy between the two systems (Table 1). By further taking the phonon vibrational energies into consideration (Figure 5a), the results of the simulations predict a spontaneous inversion between the hexagonal and tetragonal phases at 430 K, which is encouragingly close to the experimentally observed temperature range for the δ-to-α transition (290-400 K[31,32]) for real (defective and polycrystalline) samples.

Due to the lower vibrational entropy of the cubic phase and its consequently higher Gibbs free energy by comparison to the tetragonal one, the hexagonal-to-cubic phase transition turns out to be (free) energetically disfavored by comparison to the hexagonal-to-cubic one. These results reveal that the phase transition from non-perovskite to perovskite structures (δ-to-α transition) is a vibrational entropy driven hexagonal-to-tetragonal transition, in contrast to the previously assumed hexagonal-to-cubic transition.[27,45–47]

*FA dynamics and rotational entropy*
In the above discussion of the Gibbs free energy, the rotational entropies are considered by computing the so-called frustrated (rotational and translation) vibrational modes in the phonon dispersion. In a previous report claiming that the rotational entropy ($S_{rot}$) is the driving force of phase transition, $S_{rot}$ for the cubic structure was calculated as follows[32],

$$S_{rot}(T) = \frac{3}{2} k_B \{1 + \ln(0.4786\, k_B\, T\, \sqrt[3]{I_1\, I_2\, I_3})\}$$



Conversely, the rotational entropy in the hexagonal structure was taken as zero, due to the experimental neutron diffraction observation of frozen FA$^+$ dynamics in the hexagonal phase under 220 K.[32] This is substantially lower than the actual temperature range for the δ-to-α transition (290-400 K[31,32]), for which the rotational FA$^+$ dynamics may become activated.

To clarify these aspects, in the following we turn to the dynamics of the FA$^+$ cations by analyzing the underpinning potential energy surfaces and DFT-MD trajectories for the α and δ phase at 330 K. Figure 6(a-c) report the calculated rotational barriers for the FA$^+$ cation in the cubic, tetragonal, and hexagonal phases of FAPbI$_3$. Rotations of the FA$^+$ cations are modeled with respect to the three axes shown in Figure 6(d–f) and passing across the C-H bond (C-H), the two N-atoms (N-N), and the C-atom perpendicularly to the molecular plane (C-atom) of FA$^+$, respectively.

As seen in Figure 6 (a-c) and Table S2, the N-N rotations of the FA$^+$ cations have the lowest energy barriers in all three different structures. Given its dynamic nature (Figure 4), to estimate the rotational energy barriers of FA$^+$ cations in the α phase, we can assume it is bracketed by the results for the cubic and tetragonal structures i.e., 0.22-0.34 eV (Figure 6 and Table S2). For the δ phase, such a barrier is reduced (0.19 eV), which indicates facilitated rotational dynamics of the FA$^+$ cations in the δ phase.

Overall, these results inform us that the rotational entropies in the α and δ phase are similar. Analysis of the rotational angle around the N-N axis of the FA$^+$ cations in both phases during the DFT-MD trajectories (Figure 6g), confirms this statement. The rotational angle range and frequency are alike in both phases. Based on these results, it is to be concluded that the rotational entropy difference in α and δ phase cannot be the driving force of the phase transition.

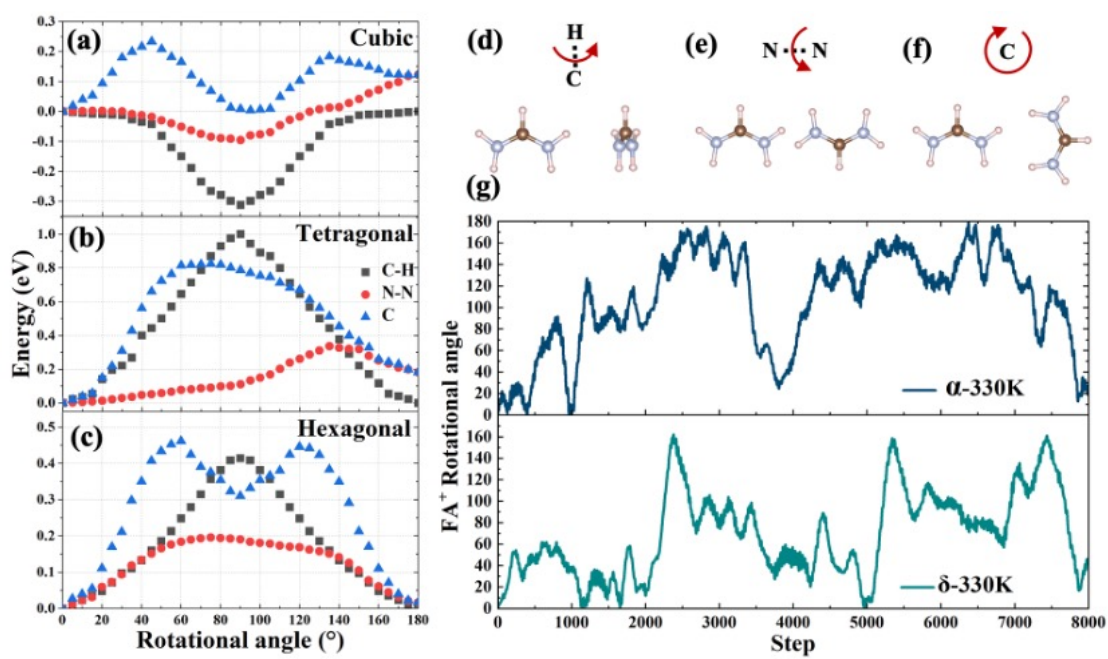

**Figure 6. Energy barriers and DFT-MD simulations of the FA$^+$ rotational dynamics in the α and δ phases.** (a-c) Rotational energy barriers for the FA$^+$ cations



in the cubic, tetragonal, and hexagonal structures as a function of the rotation angle. (d-f) schematic representations of the main FA$^+$ rotations: the arrows mark the directions of rotation. (g) The evolution of the FA$^+$ rotational angles along the N-N axis during the MD-DFT trajectory (8 ps, 330 K) for the α and δ phase.

## Conclusions

In summary, we have presented a comprehensive computational investigation aimed at unravelling the intrinsic mechanism of the δ-to-α phase transition of FAPbI$_3$. DFT geometry optimization of the cubic, tetragonal, and hexagonal structures at 0 K confirm the former (latter) as the systems of highest (lowest) internal energy. The occurrence of imaginary soft modes for the cubic structure indicates a dynamic instability for the cubic structure. Through DFT-MD simulations, it is shown the high temperature α phase is not the generally assumed static cubic structure, but a dynamically distorted tetragonal one. It is numerically demonstrated that, when sampled over sufficiently long-times, as needed for X-ray and/or neutron diffraction experiments, such a dynamically distorted tetragonal system yields a cubic time-averaged structure. On the basis of such a reassignment of the α phase to a dynamical tetragonal structure, combined DFT and lattice dynamics results reveal the δ-to-α phase transition to be driven by changes in the vibrational entropy of the hexagonal and tetragonal structures. These results shed new light on hitherto overlooked atomistic factors governing the stabilization and transitions between different phases of FAPbI$_3$ and, more generally, hybrid organic-inorganic perovskites for solar cells and opto-electronic applications.

## Methods

### DFT calculations

Periodic boundary conditions DFT simulations in the PAW formalism were carried out with the Vienna Ab Initio Simulation Package (VASP), using the Perdew, Burke, and Ernzerhof (PBE) functional.[48–50] Van der Waals (vdW) interactions were accounted for at DFT-D3 level.[39,40] We performed DFT calculations for the three (cubic, tetragonal, and hexagonal) competing room-temperature phases of FAPbI$_3$. We calculated the phonon dispersion for each phase within the harmonic approximation, to identify the atomic contributions to each phonon mode and related thermal properties. To compute total energies and atomic forces, we used a 500 eV plane wave cut-off and a 4 × 4 × 4 Monkhorst–Pack k-point grid. The valence electrons were taken into consideration as follows: C $s^2p^2$, N $s^2p^3$, H $1s^1$, Pb $6s^26p^2$, I $s^2p^5$. All structures were fully relaxed below the maximum force threshold of 0.005 eV/Å. The computational models we used for geometry optimization of the cubic and tetragonal phases are 48-atom cells ($\sqrt{2}$a × $\sqrt{2}$b



× 2c supercell expansions of the corresponding primitive cell). The lattice transformation matrix is as follows:

$$\begin{pmatrix} 1 & -1 & 0 \\ 1 & 1 & 0 \\ 0 & 0 & 2 \end{pmatrix}$$

It should be stated that with the supercell expansion, the possible orientations for the $FA^+$ cation increase, and the symmetry is lowered with respect to the primitive cell. For the hexagonal phase, we used its 48-atom primitive cell in all the calculations. The optimized lattice parameters are shown in Table 1. Further details of this structure can be found in the Supplementary Information.

*Lattice dynamics analysis*
The lattice dynamics analysis was performed by solving the dynamical matrix built from the DFT calculated force constants, using the PHNONOPY package within the harmonic approximation. We used 2 × 2 × 3 supercells with displacements for three structures. For both the cubic and tetragonal structures, 72 displacements are obtained. For the hexagonal structure, 48 displacements are obtained.

The Gibbs free energy of the system is calculated according to the quasi-harmonic approximation,

$$G = U + F_{ph}$$
$$F_{ph} = E_{vib} - TS_{vib}$$

where U is the internal energy (approximated by the DFT energy) and $F_{ph}$ is the phonon free energy. $E_{vib}$ corresponds to the vibrational energy and $S_{vib}$ refers to the vibrational entropy. The vibrational entropy $S_{vib}$ is computed from the phonon density of states as:

$$S_{Vib}(T) = 3k_B \int g(\varepsilon)\{[n(\varepsilon) + 1]\ln[n(\varepsilon) + 1] - n(\varepsilon)\ln[n(\varepsilon)]\}\, d\varepsilon$$

where $k_B$ is the Boltzmann constant, $g(\varepsilon)$ is the normalized phonon DOS with energy $\varepsilon = \hbar\omega$ ($\omega$ is the mode frequency), $n(\varepsilon)$ is Bose-Einstein population of a state of energy $\varepsilon$ at temperature T.

*DFT-MD simulations*
DFT-MD simulations were carried out with the CP2K/Quickstep package.[51] Simulations of the α phase was carried out at 250 K, 330 K, and 400 K. Simulations of the δ phase was carried out at 330 K only. The analysis of octahedral tilting and FA-rotations during the DFT-MD trajectories was performed by means of the VMD package.[52] 3 × 3 × 3 supercells for the cubic structure and 2 × 2 × 1 supercells for the hexagonal structure are used.

*Two-dimensional rigid calculations*
To calculate the rotational energy surface of the $FA^+$ cations in the cubic, tetragonal, and hexagonal structures, we generated a series of structures with rigidly rotated $FA^+$ cations in the $FAPbI_3$ cavities. Lattice parameters and atomic positions were kept fixed in the calculations. Three rotational axes were taken into consideration, which are around the C-H axis, N-N axis, and C atoms, as shown in Figure 6(d-f).

## Acknowledgements.

This work was financially supported by the National Natural Science Foundation of China (11974037，U1930402), and the Key Science and Technology Research Project of Yunnan (202002AB080001-1-6 and 202102AB0800008). G.T. and L.M.L. acknowledge support by the Royal Society Newton Advanced Fellowship scheme (grant No. NAF\R1\180242). We are grateful to the Beihang HPC and Tianhe2-JK for generous grants of computer time. We are also grateful for technical support from the High Performance Computing Center of Central South University.

## Reference

1. Saparov, B. & Mitzi, D. B. Organic-Inorganic Perovskites: Structural Versatility for Functional Materials Design. *Chemical Reviews* **116**, 4558–4596 (2016).
2. Li, W. *et al.* Chemically diverse and multifunctional hybrid organic–inorganic perovskites. *Nature Reviews Materials* **2**, 1–18 (2017).
3. Xu, W.-J., Du, Z.-Y., Zhang, W.-X. & Chen, X.-M. Structural phase transitions in perovskite compounds based on diatomic or multiatomic bridges. *CrystEngComm* **18**, 7915–7928 (2016).
4. Kieslich, G. & Goodwin, A. L. The same and not the same: molecular perovskites and their solid-state analogues. *Materials Horizons* **4**, 362–366 (2017).
5. Zhang, W. & Xiong, R. G. Ferroelectric metal-organic frameworks. *Chemical Reviews* **112**, 1163–1195 (2012).
6. Brivio, F. *et al.* Lattice dynamics and vibrational spectra of the orthorhombic, tetragonal, and cubic phases of methylammonium lead iodide. *Physical Review B - Condensed Matter and Materials Physics* **92**, 1–8 (2015).
7. Rong, Y. *et al.* Challenges for commercializing perovskite solar cells. *Science* **361**, (2018).
8. Zhou, H. *et al.* Interface engineering of highly efficient perovskite solar cells. *Science* **345**, 542–546 (2014).
9. Tong, C. J., Li, L., Liu, L. M. & Prezhdo, O. v. Long Carrier Lifetimes in PbI2-Rich Perovskites Rationalized by Ab Initio Nonadiabatic Molecular Dynamics. *ACS Energy Letters* **3**, 1868–1874 (2018).
10. Tong, C. J., Li, L., Liu, L. M. & Prezhdo, O. v. Synergy between Ion Migration and Charge Carrier Recombination in Metal-Halide Perovskites. *Journal of the American Chemical Society* **142**, 3060–3068 (2020).
11. Tong, C. J., Geng, W., Prezhdo, O. v. & Liu, L. M. Role of Methylammonium Orientation in Ion Diffusion and Current-Voltage Hysteresis in the CH3NH3PbI3 Perovskite. *ACS Energy Letters* **2**, 1997–2004 (2017).
12. Tong, C. J. *et al.* Uncovering the Veil of the Degradation in Perovskite CH3NH3PbI3 upon Humidity Exposure: A First-Principles Study. *Journal of Physical Chemistry Letters* **6**, 3289–3295 (2015).

13. Kojima, A., Teshima, K., Shirai, Y. & Miyasaka, T. Organometal halide perovskites as visible-light sensitizers for photovoltaic cells. *Journal of the American Chemical Society* **131**, 6050–6051 (2009).
14. Green, M. A., Ho-Baillie, A. & Snaith, H. J. The emergence of perovskite solar cells. *Nature Photonics 2014 8:7* **8**, 506–514 (2014).
15. Shin, S. S. *et al.* Colloidally prepared La-doped BaSnO3 electrodes for efficient, photostable perovskite solar cells. *Science* **356**, 167–171 (2017).
16. Burschka, J. *et al.* Sequential deposition as a route to high-performance perovskite-sensitized solar cells. *Nature* **499**, 316–319 (2013).
17. Yang, W. S. *et al.* Iodide management in formamidinium-lead-halide–based perovskite layers for efficient solar cells. *Science* **356**, 1376–1379 (2017).
18. Yoo, J. J. *et al.* Efficient perovskite solar cells via improved carrier management. *Nature* **590**, 587–593 (2021).
19. Han, Q. *et al.* Single Crystal Formamidinium Lead Iodide (FAPbI3): Insight into the Structural, Optical, and Electrical Properties. *Advanced Materials* **28**, 2253–2258 (2016).
20. Sutherland, B. R. & Sargent, E. H. Perovskite photonic sources. *Nature Photonics* **10**, 295–302 (2016).
21. McMeekin, D. P. *et al.* A mixed-cation lead mixed-halide perovskite absorber for tandem solar cells. *Science* **351**, 151–155 (2016).
22. Snaith, H. J. Present status and future prospects of perovskite photovoltaics. *Nature materials* **17**, 372–376 (2018).
23. Miyata, A. *et al.* Direct measurement of the exciton binding energy and effective masses for charge carriers in organic–inorganic tri-halide perovskites. *Nature Physics* **11**, 582–587 (2015).
24. Chen, B., Yang, M., Priya, S. & Zhu, K. Origin of J–V hysteresis in perovskite solar cells. *The journal of physical chemistry letters* **7**, 905–917 (2016).
25. Kanno, S., Imamura, Y. & Hada, M. Theoretical Study on Rotational Controllability of Organic Cations in Organic-Inorganic Hybrid Perovskites: Hydrogen Bonds and Halogen Substitution. *Journal of Physical Chemistry C* **121**, 26188–26195 (2017).
26. Gélvez-Rueda, M. C., Renaud, N. & Grozema, F. C. Temperature Dependent Charge Carrier Dynamics in Formamidinium Lead Iodide Perovskite. *Journal of Physical Chemistry C* **121**, 23392–23397 (2017).
27. Weller, M. T., Weber, O. J., Frost, J. M. & Walsh, A. Cubic Perovskite Structure of Black Formamidinium Lead Iodide, α-[HC(NH2)2]PbI3, at 298 K. *Journal of Physical Chemistry Letters* **6**, 3209–3212 (2015).
28. Fabini, D. H. *et al.* Reentrant Structural and Optical Properties and Large Positive Thermal Expansion in Perovskite Formamidinium Lead Iodide. *Angewandte Chemie* **128**, 15618–15622 (2016).
29. Weller, M. T., Weber, O. J., Henry, P. F., di Pumpo, A. M. & Hansen, T. C. Complete structure and cation orientation in the perovskite photovoltaic methylammonium lead iodide between 100 and 352 K. *Chemical communications* **51**, 4180–4183 (2015).
30. Fabini, D. H. *et al.* Universal dynamics of molecular reorientation in hybrid lead iodide perovskites. *Journal of the American Chemical Society* **139**, 16875–16884 (2017).
31. Ruan, S. *et al.* Incorporation of γ-butyrolactone (GBL) dramatically lowers the phase

transition temperature of formamidinium-based metal halide perovskites. *Chemical Communications* **55**, 11743–11746 (2019).

32. Chen, T. *et al.* Entropy-driven structural transition and kinetic trapping in formamidinium lead iodide perovskite. *Science Advances* **2**, 1–7 (2016).
33. Stoumpos, C. C., Malliakas, C. D. & Kanatzidis, M. G. Semiconducting tin and lead iodide perovskites with organic cations: Phase transitions, high mobilities, and near-infrared photoluminescent properties. *Inorganic Chemistry* **52**, 9019–9038 (2013).
34. Feng, H.-J., Huang, J. & Cheng Zeng, X. Photovoltaic diode effect induced by positive bias poling of organic layer-mediated interface in perovskite heterostructure α-HC(NH$_2$)$_2$PbI. doi:10.1002/admi.201600267.
35. Yang, R. X., Skelton, J. M., da Silva, E. L., Frost, J. M. & Walsh, A. Spontaneous octahedral tilting in the cubic inorganic cesium halide perovskites CsSnX3 and CsPbX3 (X = F, Cl, Br, I). *Journal of Physical Chemistry Letters* **8**, 4720–4726 (2017).
36. Bechtel, J. S. & van der Ven, A. *Octahedral Tilting Instabilities in Inorganic Halide Perovskites*.
37. Quarti, C. *et al.* Structural and optical properties of methylammonium lead iodide across the tetragonal to cubic phase transition: Implications for perovskite solar cells. *Energy and Environmental Science* **9**, 155–163 (2016).
38. Quarti, C., Mosconi, E. & de Angelis, F. Structural and electronic properties of organo-halide hybrid perovskites from ab initio molecular dynamics. *Physical Chemistry Chemical Physics* vol. 17 9394–9409 (2015).
39. Grimme, S. Accurate description of van der Waals complexes by density functional theory including empirical corrections. *Journal of Computational Chemistry* **25**, 1463–1473 (2004).
40. Grimme, S. Semiempirical GGA-type density functional constructed with a long-range dispersion correction. *Journal of Computational Chemistry* **27**, 1787–1799 (2006).
41. Grimme, S., Antony, J., Ehrlich, S. & Krieg, H. A consistent and accurate ab initio parametrization of density functional dispersion correction (DFT-D) for the 94 elements H-Pu. *The Journal of Chemical Physics* **132**, 154104 (2010).
42. Glazer, A. M. Simple ways of determining perovskite structures. *Acta Crystallographica Section A: Crystal Physics, Diffraction, Theoretical and General Crystallography* **31**, 756–762 (1975).
43. Glazer, A. M. The classification of tilted octahedra in perovskites. *Acta Crystallographica Section B Structural Crystallography and Crystal Chemistry* **28**, 3384–3392 (1972).
44. Weber, O. J. *et al.* Phase Behavior and Polymorphism of Formamidinium Lead Iodide. *Chemistry of Materials* **30**, 3768–3778 (2018).
45. Targhi, F. F., Jalili, Y. S. & Kanjouri, F. MAPbI 3 and FAPbI 3 perovskites as solar cells: Case study on structural, electrical and optical properties. *Results in Physics* **10**, 616–627 (2018).
46. Lee, J. *et al.* Formamidinium and cesium hybridization for photo- and moisture-stable perovskite solar cell. *Advanced Energy Materials* **5**, 1501310 (2015).
47. Zheng, X. *et al.* Improved Phase Stability of Formamidinium Lead Triiodide Perovskite by Strain Relaxation. *ACS Energy Letters* **1**, 1014–1020 (2016).

48. Perdew, J. P., Burke, K. & Ernzerhof, M. Generalized gradient approximation made simple. *Physical Review Letters* (1996) doi:10.1103/PhysRevLett.77.3865.
49. Kresse, G. & Furthmüller, J. Efficient iterative schemes for ab initio total-energy calculations using a plane-wave basis set. *Physical Review B - Condensed Matter and Materials Physics* (1996) doi:10.1103/PhysRevB.54.11169.
50. Kresse, G. & Furthmüller, J. Efficiency of ab-initio total energy calculations for metals and semiconductors using a plane-wave basis set. *Computational materials science* **6**, 15–50 (1996).
51. VandeVondele, J. *et al.* Quickstep: Fast and accurate density functional calculations using a mixed Gaussian and plane waves approach. *Computer Physics Communications* **167**, 103–128 (2005).
52. Humphrey, W., Dalke, A. & Schulten, K. VMD: visual molecular dynamics. *Journal of molecular graphics* **14**, 33–38 (1996).